# Peut-on former les enseignant·e·s en un rien de temps ?

Christelle Mariais[1], David Roche[2], Laurence Farhi[1], Sabrina Barnabé[3],
Sonia Cruchon[4], Sophie de Quatrebarbes[5], Thierry Viéville[1]

[1] Inria, Learning Lab et Mission de Médiation Scientifique prenom.nom@inria.fr
[2] Académie de Grenoble, Lycée Guillaume Fichet de Bonneville (Haute-Savoie).
[3] SNJazur http://snjazur.fr snjazur@gmail.com
[4] 4minutes34 https://www.4minutes34.com sonia.cruchon@gmail.com
[5] S24B http://www.s24b.com sophiedequatrebarbes@gmail.com

**Résumé (250 mots).** En France, on a récemment compris qu'il est urgent de ne plus attendre pour initier nos filles et nos garçons à l'informatique pour maîtriser le numérique : l'enseignement de Sciences Numériques et Technologie (SNT) va, en classe de seconde à la rentrée 2019, concerner tou·te·s les lycéen·ne·s. Mais comment former les enseignant·e·s ? C'est un défi qu'il faudrait relever en peu de temps. En utilisant des données quantitatives sur les formations Class´Code et en prenant du recul sur les éléments méthodologiques développés et expérimentés, nous faisons ici un retour d'expérience sur ce qui a pu marcher et, plus intéressant encore, là où nous avons buté afin de poser les limites de l'approche et montrer la nécessité d'une place plus grande pour ce type de formations.

**Mots-clés (≤ 5).** Formation des enseignant·e·s, Enseignement de l'Informatique, Pédagogie disruptive, Sciences Numériques et Technologie, Class´Code.

**Abstract.** In France, we recently realized that it is urgent to stop waiting to start teaching our girls and boys bases of computer science, to master digital technology, and the Digital Science and Technology (SNT) topic is now offered in Year 11 (classe de 2nd), starting at the 2019 school year. All highschool students are concerned. But how to train teachers? It is a challenge and it should be done in almost no time. By using quantitative data about the Class'Code training program and putting the methodological elements developed and experimented in perspective, we report here what did work and, more interesting, where we stumbled in order to set the limits of the approach and show the importance of a greater place regarding these formations.

## 1 Introduction

En France, on a récemment compris qu'il est urgent de ne plus attendre [1] pour initier nos filles et nos garçons à l'informatique pour maîtriser le numérique, et l'enseignement de Sciences Numériques et Technologie (SNT), en classe de seconde à la rentrée 2019, va concerner tou·te·s les lycéen·ne·s [2]. Avec 2 à 3 professeurs par lycée et environ 1500 lycées concernés, l'objectif est de former jusqu'à 5000 à 10000 professeur·e·s, sans que le chiffre soit bien établi [3]. Mais comment former ces enseignant·e·s ? C'est un défi qu'il faut par ailleurs relever en peu de temps.

Nous étudions ici ce qui a pu marcher et, plus intéressant encore, les problèmes rencontrés dans la mise en œuvre d'une formation en ligne, complémentaire des formations académiques, répondant à ce besoin. Il s'agit ici d'un retour d'expérience et non d'une véritable étude en sciences de l'éducation, d'une part parce que nous ne disposons pas des outils d'analyse nécessaires, et d'autre part parce qu'il serait plus correct qu'une telle étude soit faite de manière indépendante par une tierce partie.

## 2   Expression du besoin

On est ici dans le cadre d'une formation non-initiale avec un enjeu majeur : les enseignant·e·s à former sont déjà en activité et n'ont pas de disponibilité immédiate pour arrêter leurs activités professionnelles usuelles et se former. Il faut donc mettre en place un dispositif qui se mutualise avec les temps de préparation et de suivi des cours et les quelques heures disponibles pour la formation continue. Ce dispositif devra permettre une montée en compétence au fil de l'activité d'enseignement, en complément de nécessaires formations académiques plus complètes.

Afin de contribuer à la formation des futur·e·s professeur·e·s de SNT, l'objectif est de proposer, en s'appuyant sur le MOOC ICN[1] mis en place précédemment :
- S : des explications scientifiques pour commencer à maîtriser les connaissances de bases de l'informatique (découverte des notions de base du codage de l'information (ex : codage d'une image), de l'algorithmique et principes de la programmation (en amont de l'apprentissage de la programmation elle-même), et des notions sur le fonctionnement des réseaux et des machines.)
- N : des ressources de culture numérique pour s'approprier les différents sujets proposés aux élèves et faire le lien avec le monde qui nous entoure, sur les thématiques du programme : données, internet, localisation, web, photographie numérique, objets connectés, réseaux sociaux ; il s'agit ici de lier les notions fondamentales avec le quotidien numérique, de comprendre le fonctionnement de base de ces systèmes, et d'enrichir sa culture sur ces sujets par exemple avec des repères historiques.
- T : un peu de Technologie pour s'initier à la programmation et créer ses propres objets numériques ; c'est à ce niveau-là que l'on partage des activités pour manipuler les notions à travers les objets numériques du quotidien, avec ou sans faire de programmation (par exemple observer le fonctionnement d'un réseau).

On parle ici d'une formation (i) participative (que l'apprenant·e prend elle ou lui-même en main en s'appuyant sur des personnes ressources, dites facilitatrices), (ii) performative (on apprend avec les activités qui seront ré-utilisées avec les jeunes, et on dispose donc d'activités clés en main) et (iii) contaminante (les apprenant·e·s d'aujourd'hui ont vocation à devenir les formateurs de formateurs de demain).

## 3   Résultats obtenus avec la formation ICN précédente

En 2015 est créé l'enseignement d'Informatique et Création Numérique (ICN) [4]. C'est un enseignement d'exploration en informatique et en création numérique, qui est optionnel et qui a été proposé à un petit nombre d'élèves, impliquant les professeur·e·s les plus motivé·e·s. Il n'y avait volontairement pas de programme précis en matière de connaissances ou de savoir-faire, mais, selon le programme officiel, un objectif général « de comprendre la logique et les enjeux du traitement de l'information, d'acquérir une nouvelle maîtrise des logiques et concepts mis en œuvre dans le domaine du numérique, mais aussi de mobiliser le numérique comme vecteur de créativité ». Les élèves étaient invités à travailler en mode projet. Un besoin immédiat de formation des enseignant·e·s a émergé. En réponse, le MOOC ICN a donc été créé[1].

L'ensemble des statistiques que nous produisons ici proviennent d'une enquête par formulaires complétée par les participants au MOOC sur la base du volontariat. Deux formulaires ont été proposés : un questionnaire en début de cours (3094 répondants) et un questionnaire en fin de cours (340 répondants). Il va de soi que cette enquête possède un biais puisqu'on s'attend à ce que ce soit les personnes investies vraiment dans le MOOC, et probablement - de ce fait - les personnes le réussissant le mieux, qui ont répondu. Cette enquête a été présentée comme un outil pour écouter l'avis des participants, mieux les connaître et améliorer la formation au fil du temps. Les données collectées dans cette enquête sont anonymes, et, lors du lancement du MOOC ICN, nous n'avions pas prévu d'effectuer un recoupement entre les réponses aux différents questionnaires. Voici cependant ce que nous pouvons dire des données recueillies.

Le MOOC ICN a été utilisé à ce jour par plus de 10000 personnes (plus de 20000 inscrites en deux ans, dont la moitié ont utilisé au moins une partie des ressources). Dans cette population, selon les réponses au premier questionnaire, environ 34% sont des enseignants : cela indique que le MOOC n'intéresse pas uniquement les professeurs qui enseignent ICN. Il semble que 20 à 25% des enseignant·e·s d'ICN ont été touchés (mesure statistique approximative), comme détaillé Fig.1. Par ailleurs, plus de 2000 attestations de suivi ont été délivrées et le MOOC ICN présente un taux de satisfaction record : plus de 90% des répondants au dernier questionnaire évaluent à 4 ou 5 sur 5 leur satisfaction globale sur le MOOC. Des formations présentielles (y compris pour la formation des cadres de l'éducation nationale) et des temps de rencontre en ligne (peu suivis en temps réel, de l'ordre de 15 à 20 personnes, mais dont les enregistrements vidéos ont plusieurs centaines de vues ensuite) ont complété la formation en ligne.

Le point-clé de cette formation est qu'elle représente clairement une formation citoyenne à la culture scientifique et technique du numérique. L'étude des réponses au premier questionnaire de notre enquête permet de commencer à se représenter quelle population est intéressée par une telle initiative d'université citoyenne : le MOOC touche le monde francophone au-delà de la France (31,55%), majoritairement des personnes entre 25 et 55 ans (75,25%), environ la moitié (49,7%) découvrent le domaine. Au-delà des enseignant·e·s, ce sont principalement des salarié·e·s d'entreprise qui sont venus mieux comprendre les fondements du numérique ou élargir leur horizon professionnel. Parmi les enseignants, sans surprise, ceux de mathématiques (42,48%), physique-chimie (15,99%) et sciences et techniques industrielles (10,5%) sont majoritairement représentés, mais il faut noter qu'environ 30% sont des enseignants d'autres disciplines non scientifiques.

---

[1]   MOOC ICN : informatique et création numérique. http://tinyurl.com/y857b78j fun-mooc.fr

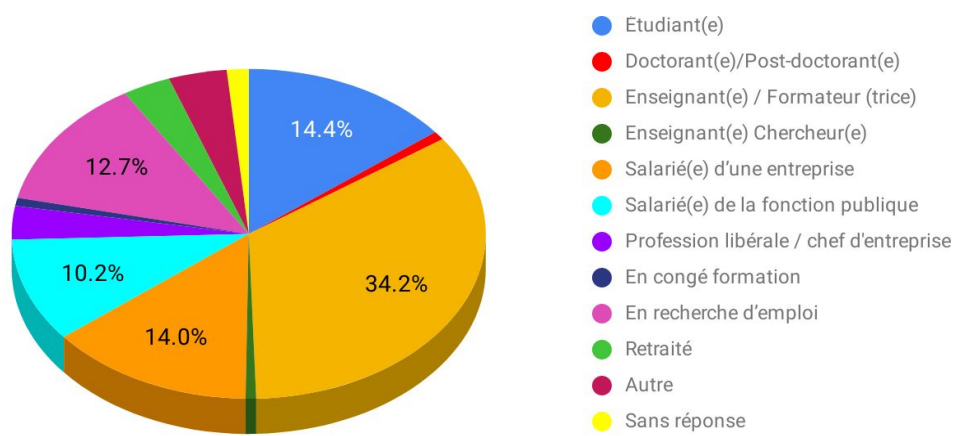

**Fig. 1.** Répartition des personnes inscrites au MOOC ICN en fonction de leur origine professionnelle, on note la proportion forte mais minoritaire d'enseignant·e·s, et la forte proportion d'étudiant·e·s (de toutes discipline) et de personnes en recherche d'emploi, utilisant cette formation comme bases pour aller vers des formations professionnalisantes.

Les évaluations sont basées sur des quiz. Les résultats des évaluations montrent que sur les 2000 personnes certifiées 20% ont plus de 85% de la note maximale. Le suivi partiel de la formation correspond souvent à des personnes qui n'ont eu besoin de suivre une partie des parcours proposés. Le taux de satisfaction très élevé correspond donc également à un taux de réussite très élevé, y compris chez les néophytes. Qualitativement, on lit beaucoup de retours du type « le cours est d'une inestimable qualité pédagogique [...] la nature hautement scientifique de son contenu, il a réussi à conjuguer simplicité, rigueur, intelligibilité … » qui, en soi, ne prouvent en rien que le cours est efficace, mais simplement qu'il est très apprécié et que les apprenant·e·s ont le sentiment qu'il est utile. Il nous semble que ce sont les bons résultats aux activités d'évaluation (quiz, rendu d'activités ou de mini-projets) qui permettent de conclure à un succès.

La disponibilité de toutes les ressources sous forme granulaire en dehors du MOOC[2] a permis une très large diffusion et la création de nouvelles ressources dont un parcours de formation

---

[2] https://classcode.fr/a-la-carte/

de Master of Science en sciences de l'éducation[3].

Nous notons que les échanges sur le forum (quelques centaines de messages) ont été en proportion très faibles et quasi exclusivement limités à des questions-réponses avec l'équipe pédagogique, majoritairement sur le fonctionnement du cours. Il semble que nous nous trouvons dans une situation de consommation passive de la formation proposée.

## 4  Rupture entre le MOOC ICN et le MOOC SNT

L'enseignement SNT est désormais un enseignement obligatoire qui va toucher tou·te·s les élèves de seconde, et donc nécessiter la formation de 5000 à 10000 professeur·e·s selon une estimation approximative. A la différence de l'ICN, c'est un enseignement avec un programme bien défini, et donc une obligation d'enseigner des notions spécifiées et de transmettre un savoir-faire en programmation de niveau débutant dans un langage de programmation précis. Le niveau théorique est du même ordre qu'en ICN, c'est le niveau d'exigence qui est augmenté.

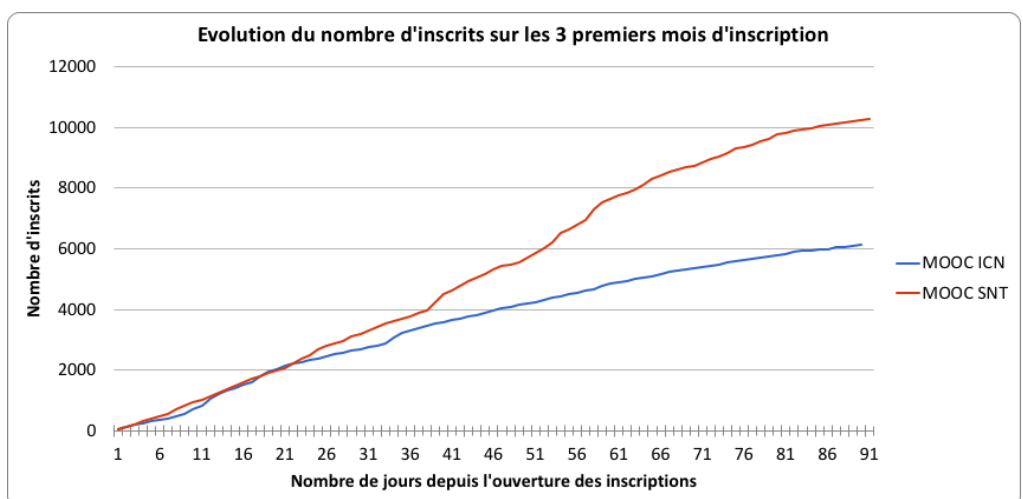

**Fig. 2.** Évolution comparative des inscriptions entre les deux MOOCs. On voit clairement une inflexion qui coïncide avec la diffusion de l'information au sein des listes de diffusion de l'Éducation Nationale.

Le 15 mars 2019, le MOOC SNT[4], adaptation du MOOC ICN complétée pour correspondre à l'enseignement SNT, est proposé. Avec plus de 8000 personnes inscrites en quelques jours, dont 25% actives sur le forum les premiers jours (≈500 posts en 8 jours, sur 200 sujets), la participation est explosive. Nous attribuons ce changement aux objectifs que se donnent les personnes qui suivent cette formation. La Fig.2. détaille cette comparaison. Dans un cadre purement culturel, acquérir une partie des compétences semble suffisant. Dans le présent

---

[3] https://classcode.fr/classcode-informatics-and-digital-creation-online-free-open-course/
[4] https://classcode.fr/snt

cadre, les participants s'imposent semble-t-il de posséder l'ensemble des éléments.

Ce MOOC offre des vidéos didactiques[5] réutilisables avec les élèves, des notebooks qui permettent de s'initier et d'initier à la programmation sans rien installer sur son ordinateur, avec des exercices de code, et des activités scolaires proposées par les collègues apprenant·e·s.

Dans ce MOOC, une enquête est également réalisée à l'aide de deux questionnaires : un questionnaire en début de MOOC (2246 réponses au 09 mai 2019) et un questionnaire de fin de cours (non exploité ici).

D'après les réponses au premier questionnaire, le MOOC SNT est majoritairement suivi par des enseignants (87%) de France (90%) dont 40% de femmes, ayant un profil disciplinaire similaire à ICN et venus se former dans la cadre de leur travail (70%), dont 31% sur proposition de l'employeur, comme l'illustre la Fig. 3, tandis que la Fig.4. donne le détail de la répartition par disciplines, des enseignant·e·s, qui reste similaire à celle en ICN.

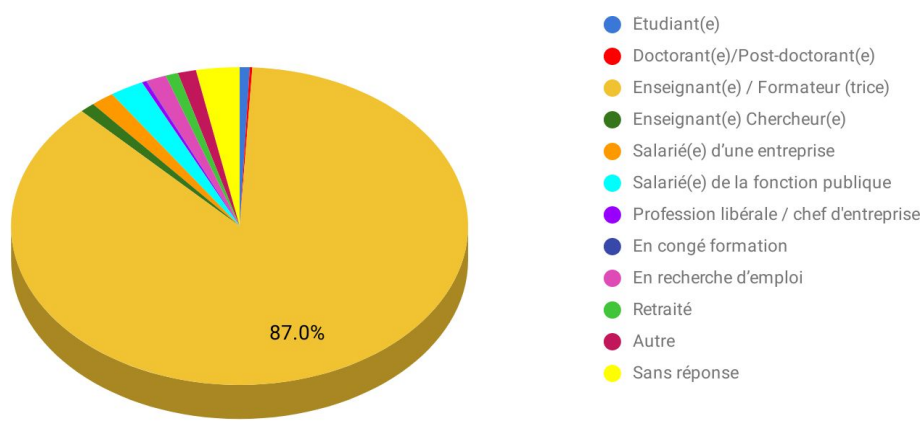

**Fig. 3.** Répartition des personnes inscrites au MOOC SNT en fonction de leur origine professionnelle, on est là clairement dans une formation spécifique pour les enseignant·e·s de cette nouvelle matière.

L'utilisation des ressources du MOOC et les premiers échanges dans le forum de discussion montrent une rupture avec la formation ICN : dans le MOOC SNT les participant·e·s sont en situation de co-acteurs ou actrices de la formation, elles et ils cherchent à avoir 100% de réponses exactes aux quiz, et questionnent sur le fond ou la façon de poser la question, parcourent les ressources avec une lecture critique et corrective, amènent des éléments complémentaires, discutent de ces ressources avec les autres participant·e·s (plus de 30% des échanges se font entre pairs). Nous sommes bien dans une situation de formation participative, performative et contaminante ; critique mais bienveillante[6].

---

[5] https://pixees.fr/classcode-v2/a-la-carte/t-thematiques-en-sciences-du-numerique
[6] exemple: «[..]vous remercier pour la qualité extraordinaire du contenu et du matériel pédagogique proposé [..] aussi pour tout le travail effectué en amont et la centralisation de l'information

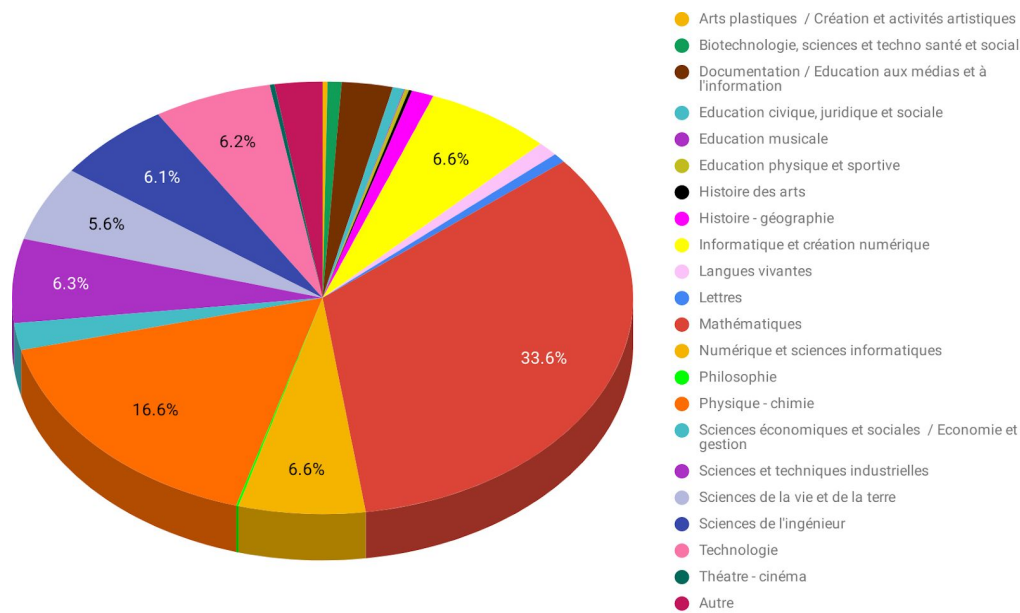

**Fig. 3.** Répartition disciplinaire des enseignant·e·s suivant la formation SNT. Cela donne aussi une vue de la répartition des professeur·e·s, par discipline, qui vont enseigner SNT au lycée. Cette proportion était qualitativement similaire pour le MOOC ICN.

On perçoit aussi cette différence entre les formations ICN et SNT dans le fait que le public actif sur le forum du MOOC SNT est majoritairement constitué d'enseignant·es : la raison à cela est qu'il y a une obligation pour les enseignants d'atteindre un niveau de connaissances théoriques minimum afin, contrairement à certaines idées reçues, de se sentir suffisamment à l'aise pour enseigner SNT. Faire acquérir ce niveau théorique minimum est, selon nous, le véritable défi dans ce MOOC et au niveau des formations académiques. La diversité des profils d'enseignants à former et l'hétérogénéité des connaissances de départ est aussi une difficulté. Bien sûr, il ne faut pas négliger l'aspect didactique, mais cet aspect didactique doit s'appuyer sur des connaissances théoriques relativement solides, comme dans le cas des autres disciplines.

## 5    Conclusion

Des formations comme les MOOC SNT et ICN semblent jouer un vrai rôle de support et d'accompagnement dans cette situation d'urgence, sans remplacer une vraie formation à la discipline informatique. De plus, ce besoin de formation dépasse largement cet enseignement

---

[..] on fait aussi une véritable économie de recherche sur les sujets présentés [..] j'espère que les enseignants de demain pourront toujours s'appuyer sur les chercheurs et les enseignants plus expérimentés durant leur carrières.».

scolaire et est une nécessité pour toute notre société.

D'une part, ces formations nous apprennent qu'il y a beaucoup d'enseignant·e·s prêt·e·s à s'investir, même s'ils sont totalement néophytes ou peu technophiles, à condition qu'une formation solide et qu'un accompagnement soient proposés. Cela brise une idée reçue, mais cela impose qu'au-delà de ce MOOC de vraies formations pluriannuelles soient proposées en formation continue et en formation initiale.

La mise en oeuvre de ces MOOC nous montre d'autre part qu'il est très important de proposer une formation de niveau vraiment introductif avant d'envisager d'aller plus loin. Cela a une conséquence fondamentale, au niveau des élèves : il faudra privilégier la qualité de ce qui est appris, et non l'exhaustivité. Aucun élève - sauf exception - n'apprendra la totalité du programme, mais que toutes et tous bénéficient d'une vraie formation à ces sujets peut changer notre société.